# Hybrid Integration of Solid-State Quantum Emitters on a Silicon Photonic Chip

*Je-Hyung Kim,[†,‡,⊥] Shahriar Aghaeimeibodi,[†,⊥] Christopher J. K. Richardson,[§] Richard P. Leavitt,[§] Dirk Englund,[∥] and Edo Waks\*,[†,#]*

[†]Department of Electrical and Computer Engineering and Institute for Research in Electronics and Applied Physics, University of Maryland, College Park, Maryland 20742, United States

[‡]Department of Physics, Ulsan National Institute of Science and Technology (UNIST), Ulsan 44919, Republic of Korea,

[§]Laboratory for Physical Sciences, University of Maryland, College Park, Maryland 20740, United States

[∥]Department of Electrical Engineering and Computer Science, Massachusetts Institute of Technology, Cambridge, Massachusetts 02139, United States

[#]Joint Quantum Institute, University of Maryland and the National Institute of Standards and Technology, College Park, Maryland 20742, United States

KEYWORDS. Silicon quantum photonics, hybrid integration, quantum dots, single photons

**Scalable quantum photonic systems require efficient single photon sources coupled to integrated photonic devices. Solid-state quantum emitters can generate single photons with high efficiency, while silicon photonic circuits can manipulate them in an integrated device structure. Combining these two material platforms could, therefore, significantly increase the complexity of integrated quantum photonic devices. Here, we demonstrate hybrid integration of solid-state quantum emitters to a silicon photonic device. We develop a pick-and-place technique that can position epitaxially grown InAs/InP quantum dots emitting at telecom wavelengths on a silicon photonic chip deterministically with nanoscale precision. We employ an adiabatic tapering approach to transfer the emission from the quantum dots to the waveguide with high efficiency. We also incorporate an on-chip silicon-photonic beamsplitter to perform a Hanbury-Brown and Twiss measurement. Our approach could enable integration of pre-characterized III-V quantum photonic devices into large-scale photonic structures to enable complex devices composed of many emitters and photons.**

Photonic quantum information processors use multiple interacting photons to implement quantum computers,[1,2] simulators,[3,4] and networks.[5-8] These applications require efficient single photon sources coupled to photonic circuits that implement qubit interactions to create highly connected multi-qubit systems.[8-11] Scalable photonic quantum information processors require methods to integrate single photon sources with compact photonic devices that can combine many optical components. Such integration could enable complex quantum information processors in a compact solid-state material.[12-14]

Silicon has many advantages as a material for integrated quantum photonic devices. It has a large refractive index that enables many photonic components to fit into a small device size.[15-17] Electrical contacts incorporated into the photonic structure can rapidly modulate and reconfigure these components by free-carrier injection on fast timescales.[18,19] Silicon is also compatible with standard CMOS fabrication methods that can combine electronics with photonics on a large scale.[15,16] For these reasons, silicon photonics can achieve the most complex integrated photonic structures to date composed of thousands of optical components in a single chip.[20,21] However, since silicon is also an indirect bandgap material with poor optical emission properties, it has not been possible to integrate efficient atom-like quantum emitters. The most common approach is to exploit the third-order nonlinearity to create entangled photon pairs by down-conversion.[22,23] But these sources only generate heralded single photons, and extending them to an on-demand source by multiplexing remains a significant challenge. Another approach is to develop quantum dots based on Si/Ge heterostructures.[24] But these dots currently emit with very poor efficiency, and to-date there are no reports of single photon emission from them validated by photon correlation measurements. The incorporating on-demand single photon sources on a silicon photonic chip remains a difficult challenge.

In this letter, we demonstrate the integration of silicon photonics with a solid-state single photon emitter. We use a hybrid approach that combines silicon photonic waveguides with InAs/InP quantum dots that act as efficient sources of single photons at telecom wavelength.[25-27] A pick-and-place technique allows transferring of tapered InP nanobeams containing InAs quantum dots onto a silicon waveguide with nanometer-scale precision. The tapered nanobeams efficiently couple the emission from the quantum dot to the silicon waveguide. The fabricated devices exhibit clear single-photon emission, which we validate via photon correlation measurements using an on-chip silicon photonic beamsplitter. Our approach could enable deterministic fabrication of complex circuits composed of multiple single photon emitters coupled to large-scale silicon photonic devices.

Figure 1a shows a schematic of the heterogeneous integration by placement of a thin InAs/InP nanobeam on top of a silicon ridge waveguide. The quantum dots have an emission wavelength around 1300 nm, as shown in the photoluminescence spectrum (Figure 1b) measured in a bulk sample at 4K. This wavelength is well below the bandgap of silicon, ensuring that the emitted photons will experience low absorption losses. The InP nanobeam has a width of 500 nm and a thickness of 280 nm while a silicon waveguide has a width of 400 nm and a thickness of 220 nm on the 3-µm thick $SiO_2$ layer on top of a silicon. We taper the nanobeam with a tapered angle of 6º and a tapered length of 5 µm in order to adiabatically convert the photonic mode from the InP beam to the silicon waveguide.

To estimate the coupling efficiency between the nanobeam and waveguide, we simulate the mode propagation in the integrated structure using a finite-difference time-domain numerical simulation. We approximate the quantum dot emission with a dipole source located at the center of the nanobeam. Figure 1c shows the amplitude of the electric field generated by the dipole as it propagates through the tapered region. Near the center of the beam, the emission from the dipole couples to both the InP nanobeam and the Si waveguide due to their similar refractive indices of 3.2 and 3.5, respectively. But as the nanobeam tapers, the field adiabatically transforms to the single mode of the silicon waveguide, as shown in both the longitudinal cross-section in Figure 1c and the transverse mode profiles in figure 1d taken at several positions along the taper. In the simulation, we assume that the quantum dot emission propagates only one direction and calculate a 32% coupling efficiency from the dipole to the silicon waveguide mode while a non-tapered nanobeam has 13% coupling efficiency due to scattering and back-reflection at the boundary (See Method and Figure S1 in Supporting Information).

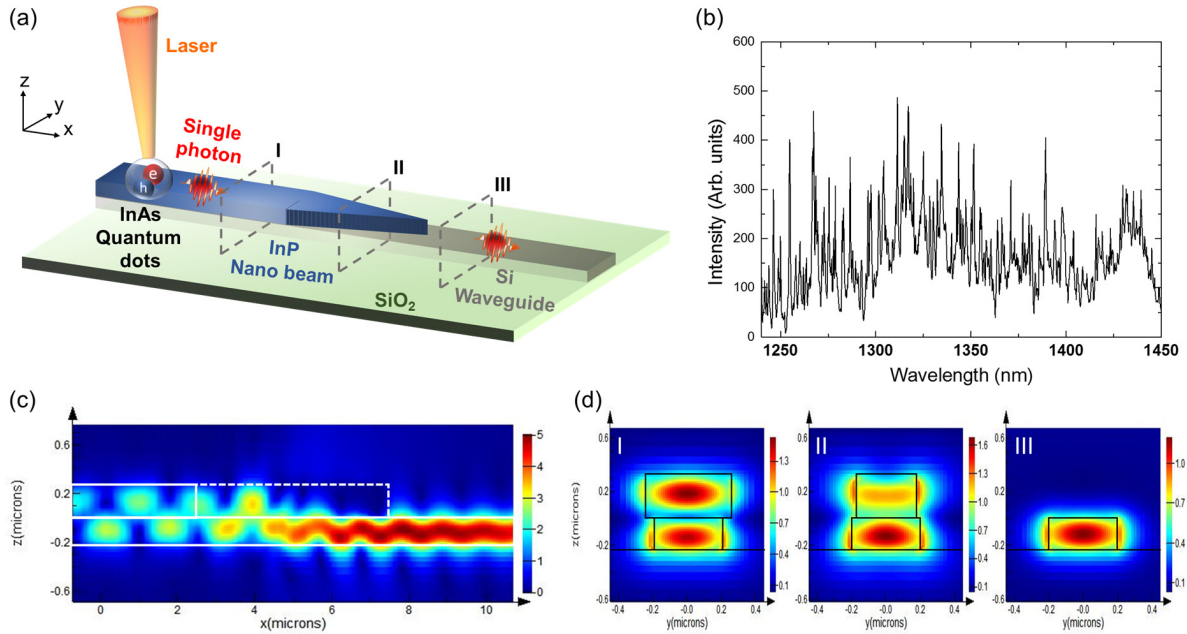

**Figure 1.** (a) Schematic of the integrated InP nanobeam and silicon waveguide. (b) Photoluminescence spectrum of the InAs quantum dots from a bulk sample. (c) Finite-difference time-domain simulation of field ($|E|$) profile at side. White-solid lines represent the boundary of the structures, and a white-dashed line indicates the tapered region in the nanobeam. (d) Simulated field ($|E|$) profiles at different cross-sectional positions of I, II, and III marked in (a). Black lines represent the boundary of the structures.

We fabricate the InP nanobeam and silicon waveguide separately using electron beam lithography and dry etching (see Methods). Figure 2a shows a scanning electron microscope image of the fabricated nanobeam which is suspended by thin tethers that attach it to the bulk substrate. The nanobeam contains a periodic array of air holes on one end that acts as a Bragg mirror to direct the quantum dot emission in only one direction. Figure 2b shows a scanning electron microscope image of a fabricated silicon photonic waveguide. A grating coupler on one end of the device couples light from the waveguide to the out-of-plane for detection. We also fabricate a y-shaped waveguide acting as a 50/50 on-chip beamsplitter as shown in Figure 2c. The square pad on the left end of the nanobeam and waveguide facilitates picking and placing the nanobeams as described below.

To integrate the fabricated InP nanobeams with the silicon waveguides, we pick the InP nanobeam with a microprobe tip and place it on the silicon waveguide. Unlike previously reported hybrid integration methods that achieved alignment using an optical microscopy[13,28,29] or a wafer bonding technique,[30] our pick-and-place technique uses a combined focused ion beam and scanning electron microscope. The focused ion beam detaches the InP nanobeam from the substrate by cutting the tethers. We locally etch the tethers placed at a sufficiently large distance away from the nanobeam to avoid damaging the quantum dots by high energy ion bombardment during the release process (inset in Figure 2d). After releasing the beam, we pick it up with a microprobe tip as shown in Figure 2d. The nanobeam adheres to the tip by Van der Waals forces. We then move the nanobeam to the silicon chip (Figure 2e) using a moveable stage and place the nanobeam on the silicon waveguide devices. We use the scanning electron microscope to image the beam as we place it on the silicon substrate, which enables us to achieve nanometer alignment accuracy with high repeatability. Figure 2f shows a scanning electron microscope image of four completed devices composed of an integrated nanobeams on a silicon waveguide. The fact that we can integrate a nanobeam on each structure in the 4x4 array of waveguides demonstrates the repeatability and high yield of this procedure.

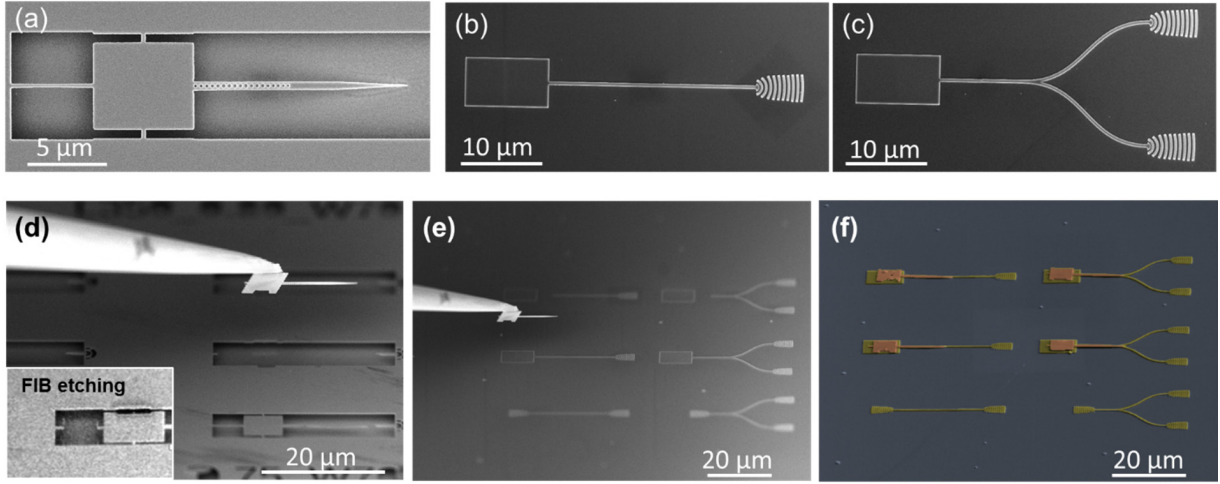

**Figure 2.** (a) Scanning electron microscope image of the fabricated tapered nanobeam containing InAs quantum dots. (b,c) Scanning electron microscope images of the fabricated silicon straight waveguide (b) and y-shaped 50:50 waveguide beam splitter (c). Square pads at the left end of the nanobeam and waveguide devices are for picking and placing the nanobeam. Gratings at the right end of the waveguides are for extracting photons from the waveguide to out-of-plane. (d-f) Pick-and-place procedure with a microprobe tip combined with a focused ion beam and scanning electron microscope. (d) A microprobe tip picks the nanobeam from the original InP template. Inset shows a locally cut tether by a focused ion beam for releasing the nanobeam. (e) Nanobeam transfer onto a silicon chip. (f) False color scanning electron microscope image of the integrated nanobeam and silicon waveguide devices. Red and yellow colors indicate the nanobeam and silicon waveguide structures, respectively.

We use a low temperature (4 K) micro-photoluminescence setup to characterize the fabricated devices (see Methods). We excite the quantum dots using a 785 nm continuous wave laser and collect the signal from the grating out-coupler. Figure 3a shows the resulting photoluminescence spectrum obtained from the straight waveguide device. We observe multiple sharp peaks corresponding to single quantum dots, demonstrating that the emission from the quantum dots coupled to the waveguide mode. The quantum dot emissions from the grating show a factor of 4 increase in intensity as compared to that of the quantum dots in a bulk sample shown in Figure 1b, experimentally confirming our numerical predictions.

Using the intensity of the quantum dot labeled dot A in Figure 3a, we estimate the collection efficiency using a pulsed laser excitation (see Methods) to be 3.2±0.5% at the first lens. This efficiency is lower than the expected ideal efficiency of 11.8% including the coupling from the quantum dots to the silicon waveguide and the out-coupling efficiency from the grating (see Method), which may be due to imperfect alignment of the beam with the waveguide, scattering due to fabrication disorder, quantum dot dipole orientation, and imperfect reflections from the Bragg mirror.

Next, we investigate a device that integrates the nanobeam with an on-chip beam splitter. Figure 3b plots the collected photoluminescence from each grating outcoupler when we excite the dots near the center of the nanobeam. Each spectrum shows a number of resonant peaks corresponding to different quantum dots. We identify 16 independent resonant peaks, which we label in the figure. Each peak appears in both spectra, verifying that they originate from the same source. We independently measure the splitting ratio of the beamsplitter by sending a 1304 nm laser into the waveguide and comparing the intensities at the two gratings, giving a splitting ratio of 43:57 (See Figure S3 in Supporting Information).

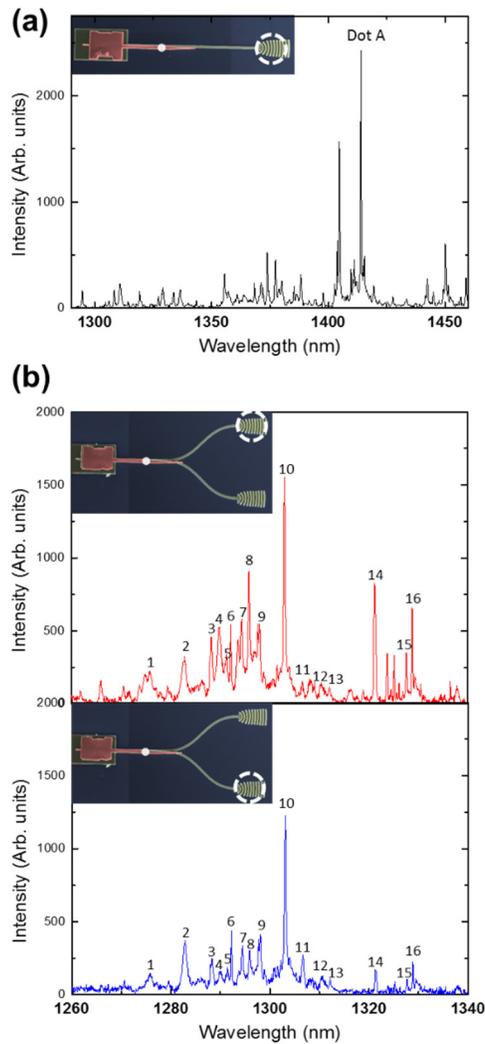

**Figure 3.** (a) Photoluminescence spectrum of the quantum dots on a straight silicon waveguide. Inset: False color scanning electron microscope image of the integrated nanobeam on a waveguide. White spot and dashed circle represent the position of laser excitation and emission collection. (b) Photoluminescence spectrum of the quantum dots on a 50/50 waveguide beam splitter. Top and bottom spectra represent the measured signal from top and bottom grating outcouplers (white dashed circle in insets) by focusing the laser at the center of the nanobeam (white dot in insets). Numbers indicate the resonant quantum dot lines on each spectrum from different gratings based on their spectral position.

The beamsplitter structure allows direct measurements of second-order correlation to confirm the single photon nature of the quantum dot emission. We send the collected light from each grating coupler to separated spectrometers using a pickoff mirror in order to filter the quantum dot line, and then couple the filtered signal to a single photon detector. Figure 4 shows the measured second-order correlation function $g^2(\tau)$ from the quantum dot line 10 using the 50/50 waveguide beam splitter after detector dark count subtraction. Fitting the antibunching dip to an equation of the form $g^2(\tau) = 1-(1- g^2(0))\exp(-|\tau|/\tau_0)$ and convolving with a Gaussian to account for time resolution (200 ps) of our system produces $g^2(0) = 0.33$, which is below the classical limit of 0.5. We attribute the residual multi-photon events to residual background emission due to above-band pumping. This contribution could be fully eliminated using resonant[31-33] or quansi- resonant[34,35] excitation. From the fit we determine the lifetime to be $\tau_0 = 0.9$ ns, which agrees with the lifetime of 1.25 ns determined from time-resolved lifetime measurements (see Figure S4 in Supporting Information).

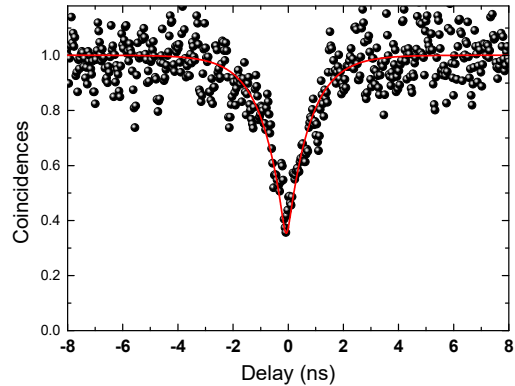

**Figure 4.** Second-order correlation function of the single quantum dot emission (line 10 in Figure 3b) from two grating outcouplers. Red solid-line is a fitted curve for $g^2(\tau)$.

In summary, we have demonstrated deterministic integration of solid-state quantum emitters on a silicon photonic device. This integration opens up the possibility to leverage the highly-advanced photonic capabilities developed in silicon to control and route non-classical light from on-demand single photon sources. In addition, the fabricated devices operate at telecom wavelengths which are useful for fiber-based quantum communication. Our technique could also solve the problems of spatial and spectral randomness in quantum dots which severely degrade device yield and serve as one of the main challenges for scalability. For example, by pre-characterizing fabricated quantum dot devices to select and then transfer only the ones that contain a dot with the desired properties, we could combine multiple quantum emitters with a desired wavelength on the same chip. Local tuning methods based on temperature, strain, and quantum confined Stark effect could provide additional fine-tuning to compensate for small residual spectral mismatch as well as to control on-chip interactions.[36-38] In the current device, the grating couplers are the most significant source of loss. Improved grating couplers using partial etching can achieve greater than 90% efficiency.[39] Alternatively, edge coupling,[40] tapered fiber coupling,[41] or integrating detectors directly on a chip[42,43] would significantly improve the overall efficiency. Ultimately, our results represent an important step towards complex quantum photonic circuits that could process many photons on a chip to simulate complex chemical reactions,[3,44,45] attain Heisenberg-limited interferometric phase sensitivity,[46,47] and implement photonic quantum computation.[1,8,48]

**Methods.**

*Sample information*. We grew the quantum dots using molecular beam epitaxy. InAs quantum dots have a density of approximately 10 $\mu m^{-2}$ in a 280 nm-thick InP membrane on a 2 $\mu$m-thick AlInAs sacrificial layer. We patterned the nanobeam device using electron beam lithography followed by inductively coupled plasma reactive ion etching and used a chemical wet etch ($H_2O$:HCl:$H_2O_2$=3:1:1) to remove the sacrificial layer leaving a suspended beam supported by several thin tethers. The nanobeam has a width of 500 nm and tapered length of 5 μm and includes a Bragg mirror consisting of air hole arrays with a radius of 100 nm and pitch of 350 nm. We fabricated silicon waveguides on a 220 nm-thick silicon on 3 μm-thick $SiO_2$ layer using electron beam lithography combined with metal liftoff. The silicon waveguides have a width of 400 nm, and the grating outcoupler has a period of 550 nm and 50% duty cycle.

*Experimental set-up.* For optical characterization, we mounted the sample on a low-vibration closed cycle cryostat operating at a base temperature of 4 K. We used a 780 nm continuous wave laser to excite the quantum dots and collect the signal from the grating outcoupler for the integrated devices using an objective lens (NA=0.7, 100x) and then sent the signal to a spectrometer for spectrum analysis and spectral filtering. For the second-order correlation measurements, we individually collected the signal from each grating by separating them using a pickoff mirror and send to spectrometers followed by fiber-coupled InGaAs single photon detectors and a time-correlated single photon counter.

*Estimation of the collection efficiency.* We calculated the collection efficiency of the emission quantum dots at the first lens (NA=0.7) using a finite-difference time-domain simulation. A dipole source at the center of the nanobeam mimics the quantum dots in the simulation. From this simulation, we calculated the coupling efficiency of 71% from the quantum dots to the integrated nanobeam and waveguide structure, the coupling efficiency of 45% from the integrated structures to the silicon waveguide, and the out-coupling efficiency of 37% at the grating coupler, resulting in a total collection efficiency of 11.8% at the first lens. In the simulation, we assumed that the quantum dot emission propagates only one direction.

*Measurement of the collection efficiency.* To estimate the brightness of the waveguide-coupled quantum dots, we excited the quantum dots with a 40 MHz pulsed laser and measured the photon counting rate of 10.8 kHz at the single photon detector. Based on our spectroscopy system efficiency of 0.85%, including a transmission efficiency of optics (35%) and spectrometer (38%), fiber coupling efficiency to the detector (32%), and the detector quantum efficiency (20%), we determined a collection efficiency of 3.2% at the first lens for the measured quantum dot A in Figure 3a.

## ASSOCIATED CONTENT

**Supporting Information**.

The Supporting Information is available free of charge via the Internet at http://pubs.acs.org. Additional finite-difference time-domain simulation for the non-tapered nanostructure, grating coupler, measurement of a splitting ratio of 50/50 beamsplitter, lifetime of quantum dots.

## AUTHOR INFORMATION

**Corresponding Author**

*E-mail: edowaks@umd.edu

**Author Contributions**

[⊥] These authors contributed equally to this work.

## ACKNOWLEDGMENT

The authors would like to acknowledge support from the Laboratory for Telecommunication Sciences, The Center for Distributed Quantum Information at the University of Maryland and Army Research Laboratory, and the Physics Frontier Center at the Joint Quantum Institute.

# Supporting Information

1. **Coupling efficiency for tapered and non-tapered nanobeams**

We simulate the mode propagation in the integrated InP nanobeam and silicon waveguide devices. Figure S1a-d(e-h) show the light propagation along the integrated structure for the non-tapered (tapered) nanobeam. InP and silicon have similar refractive indices of 3.3 and 3.5, respectively and therefore, the emission from the quantum dots in the nanobeam couple to both nanobeam and waveguide during the propagation. The mode in the non-tapered InP nanobeam meets a sudden change of refractive index at the boundary, resulting in scattering (Figure S1c) and back reflection (Figure S1d) at the boundary, while the mode in the tapered nanobeam adiabetically couple to the silicon waveguide. We calcuate the coupling efficiency of 32% for the tapered nanobeam, higher than the coupling efficiency of 13% for the non-tapered nanobeam.

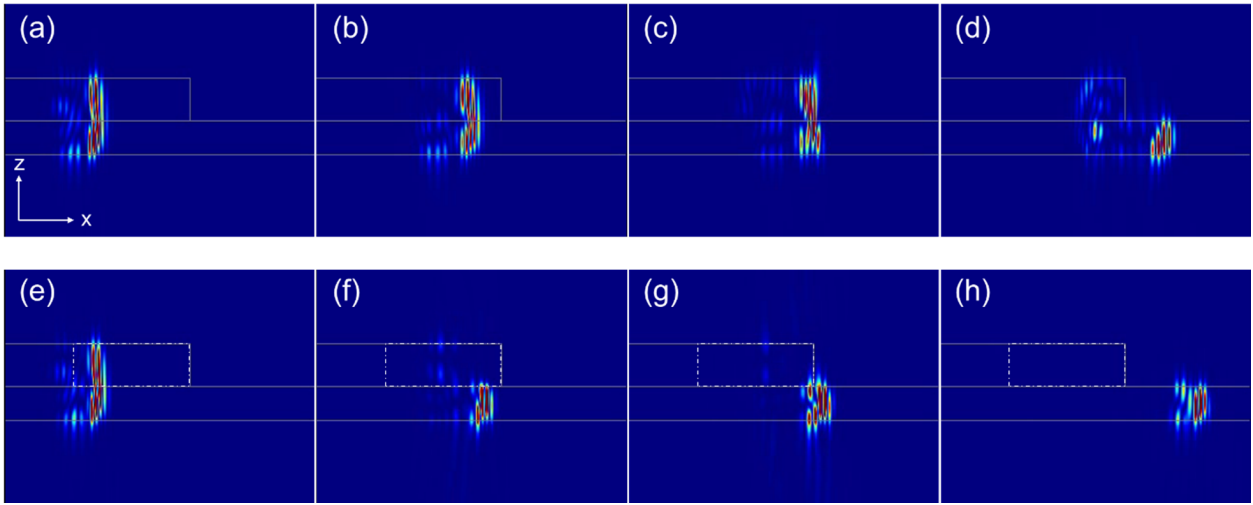

Figure S1. (a-h) Mode propagation along the integrated nanobeam and waveguide with time for non-tapered (a-d) and tapered nanobeams (e-h). White-solid lines indicate the boundary of the simulated structure, and white-dashed lines indicate the tapered region of the nanobeam.

2. **Coupling efficiency of the grating coupler**

Our grating couplers consist of a funnel-like circular sector and ten curved teeth located peridiocally at the end the waveguide (See Figure S2a). The grating has a period of 550 nm and duty cycle of 50%, and outer radius of the ciruclar sector of 500 nm, optimized using a finite-difference time-domain numerical simulation. The grating coupler shows an out-coupling efficiency of 36.8% for TE mode and 3% for TM mode of the waveguide with a N.A.=0.7 objective lens. Figure S2b shows a scanning electron microscope image of the fabricated grating.

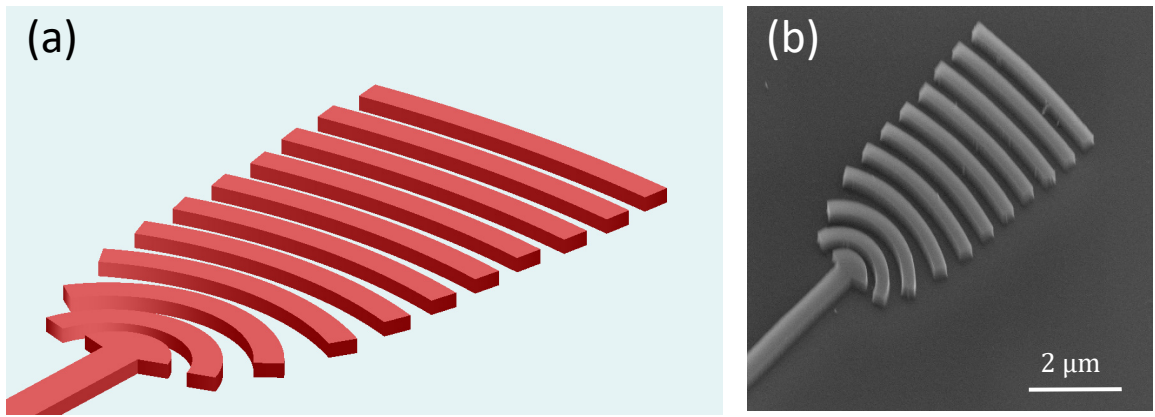

Figure S2. (a) Design of the grating coupler. (b) Scanning electron microscope image of the fabricated grating.

3. On-chip waveguide beamsplitter

We measure a splitting ratio of the silicon waveguide beamsplitter by focusing a laser on an input grating and measuring the intensities from each output grating. Figure S3a shows a scanning electron microscpe image of the fabricated 50/50 on-chip beamsplitter. Figure S3b shows the separately collected laser signal from the top and bottom gratings, and we acheive a 43:57 splitting ratio from the waveguide beamsplitter.

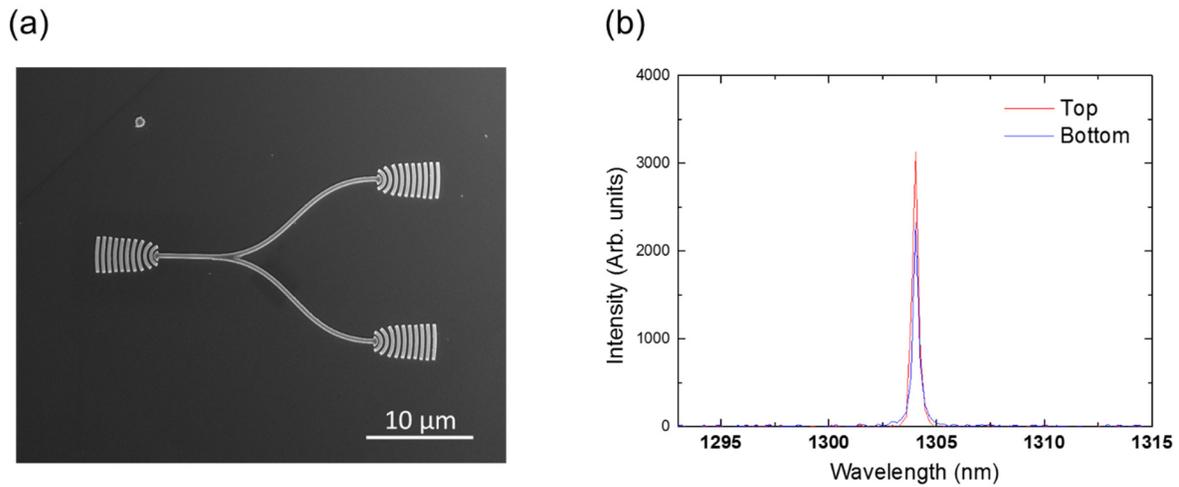

Figure S3. (a) Scanning electron microscope image of on-chip silicon beamsplitter with input and output gratings. (b) Collected laser signals from top and bottom grating couplers.

## 4. Lifetime of the quantum dots in the integrated devices.

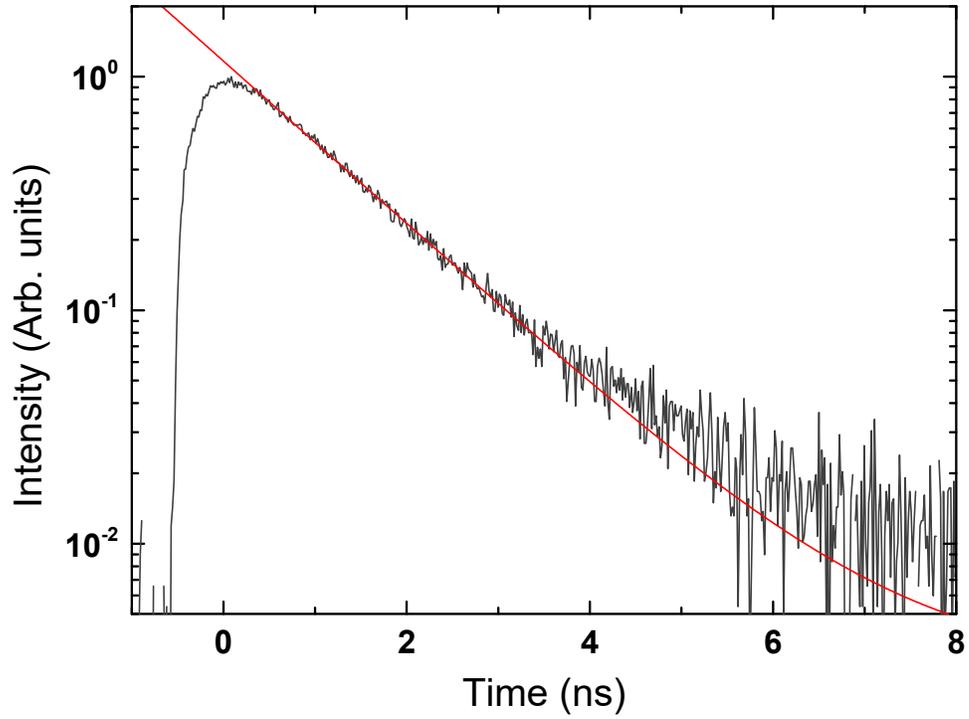

Figure S4. Decay curve for the quantum dot 10 in Figure 3b. Red lines is a fitted curve with single decay. From the fit, we determine the lifetime of 1.25±0.02 ns.